# Second harmonic generation at a time-varying interface

Romain Tirole[1], Stefano Vezzoli[1], Dhruv Saxena[1], Shu Yang[1], T.V. Raziman[1], Emanuele Galiffi[2], Stefan A. Maier[1,3], John B. Pendry[1], Riccardo Sapienza[1]

[1] The Blackett Laboratory, Department of Physics, Imperial College London; London SW7 2BW, United Kingdom

[2] Photonics Initiative, Advanced Science Research Center, City University of New York; 85 St. Nicholas Terrace, 10031, New York, NY, USA

[3] School of Physics and Astronomy, Monash University; Clayton Victoria 3800, Australia

Corresponding authors: Romain Tirole romain.tirole16@imperial.ac.uk Riccardo Sapienza r.sapienza@imperial.ac.uk

**Abstract**

Time-varying metamaterials rely on large and fast changes of the linear permittivity. Beyond the linear terms, however, the effect of a non-perturbative modulation of the medium on harmonic generation and the associated nonlinear susceptibilities remains largely unexplored. In this work, we study second harmonic generation at an optically pumped time-varying interface between air and a 310 nm Indium Tin Oxide film. We observe an enhancement of the modulation contrast at the second harmonic wavelength, up to 93% for a pump intensity of 100 GW/cm$^2$, leading to large frequency broadening and shift. We demonstrate that, in addition to the quadratic dependence on the fundamental field, a significant contribution to the enhancement comes from the temporal modulation of the second order nonlinear susceptibility, whose relative change is double that of the linear term. Moreover, the spectra resulting from single and double-slit time diffraction show significantly enhanced frequency shift, broadening and modulation depth, when compared to the infrared fundamental beam, and could be exploited for optical computing and sensing. Enhanced time-varying effects on the harmonic signal extends the application of materials to the visible range and calls for further theoretical exploration of non-perturbative nonlinear optics.

**Main**

**Introduction**

Time-varying metamaterials have both enabled novel wave phenomena and provided new perspectives on classic physical problems [1,2]. Time-varying photonics, where the linear susceptibility $\chi^{(1)}$ of the medium is modulated at optical frequencies, enables ultrafast switching of transmittivity or reflectivity (on the fs time scale) [3–5], frequency shifting and spectral modulation [6–8], beam-steering [9–11] and non-reciprocal devices [12]. Furthermore, concepts such as time-crystals [13], coherent wave control[14], and lasing [15] have been predicted and hold promise for experimental implementations.

Epsilon-near-zero materials, and more particularly transparent conductive oxides [3], have emerged as promising platforms for time-modulation at near-infrared optical frequencies [16], by combining order-of-unity changes of their linear permittivity with an ultrafast response, close to a single optical cycle [17,18]. In Indium Tin-Oxide (ITO), such changes of the refractive index at these ultrafast time scales can be described by non-perturbative photocarrier excitation [19–21]. This has led to a surge of time-varying experiments, among which are the demonstrations of time refraction [22,23], single and double slit diffraction [5,17] and single-cycle dynamics [18].

Other nonlinear effects such as harmonic generation, traditionally described by perturbative changes in the medium polarization, have also been observed in time-varying media, and frequency shift of harmonic light in high-index Silicon metasurfaces [24] and high harmonics generated in a CdO thin-film [25] have been reported. Four-wave-mixing has also been shown to undergo time refraction [26,27] and diffraction[5] in pump-probe experiments with ITO and AZO. Yet, the effects of the non-perturbative time modulation on the nonlinear susceptibility, $\chi^{(n)}$, and the nonlinear polarization $P^{(n)}$, remain mostly unexplored.

Here, we study second harmonic generation (SHG) from a time-varying interface between an ITO thin film and air, undergoing a non-perturbative time modulation, and compare it to the modulation at the fundamental frequency. We demonstrate that the optical modulation of the linear susceptibility in ITO has a large effect on the second-order nonlinear susceptibility. When accounting also for the quadratic dependence of the nonlinear polarization on the fundamental field, our simple model, based on an anharmonic oscillator, reproduces well the modulation contrast. We also show that the visible SHG signal experiences the same process of time-diffraction, including double-slit time diffraction, as the infrared beam, but with significantly enhanced frequency broadening, shift and modulation contrast. However, the observed generation of new frequencies cannot be accounted by our model and calls for further theoretical investigation of non-perturbative nonlinear optics. Our results extend time-varying studies to the visible range and open a path to new applications, like neuromorphic optical computing, that require highly nonlinear systems.

**Concept and Setup**

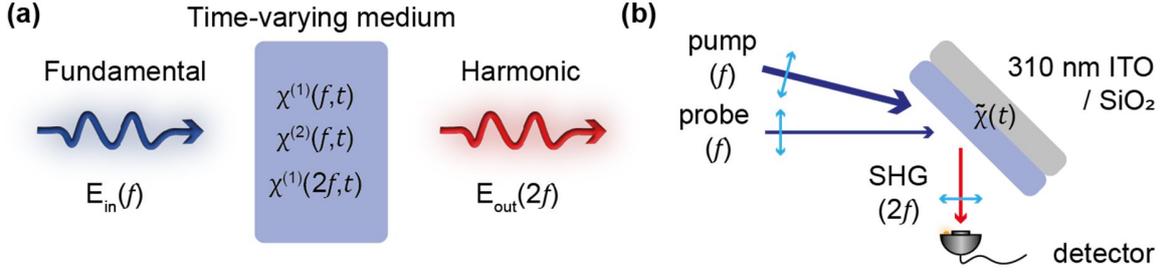

**Figure 1.** Effects of optical modulation on SHG. **(a)** Mechanisms for modulation of harmonic generated light. When a wave impinges on a time-varying medium, modulation in time of the linear susceptibility $\chi^{(1)}$ at the fundamental and second harmonic frequencies ($f$ and $2f$) as well as modulation of the nonlinear susceptibility $\chi^{(2)}$ can affect the generated light. **(b)** Diagram of the experiment: light from a probe at the fundamental frequency of 230 THz (1300 nm) generates SHG at 460 THz (650 nm) at the surface of the ITO layer, which is modulated by pumping the medium with another strong pulse at 230 THz.

Standard parametric nonlinear processes, e.g. second harmonic generation, are described in a perturbative framework by introducing a second order polarisation $P^{(2)}(2f)$ as a perturbative correction to the linear polarisation, which is the tensor product of the square of the fundamental field $E(f)$ at frequency $f$ and the second order nonlinear susceptibility $\chi^{(2)} = \chi^{(2)}(2f)$ :

$$P^{(2)}(2f) \propto \chi^{(2)} E^2(f).$$

A large modulation of the linear susceptibility $\chi^{(1)}(f)$ in a time-varying medium, due to a change of the material properties typically related to photocarrier excitation in ITO [20], has a predictable effect on the field $E(f)$ which is at the basis of time-varying metamaterials[1]. Because of the quadratic dependence of the polarisation on the fundamental field $E(f)$, one should expect an enhancement of the time-varying effects on the SHG. Moreover, the temporal modulation of $\chi^{(1)}(f)$ should also induce changes in the nonlinear susceptibility $\chi^{(2)}(2f)$ and linear susceptibility at the harmonic wavelength $\chi^{(1)}(2f)$, as sketched in Fig. 1(a). In the perturbative framework of nonlinear optics, these changes are normally considered negligible, compared to the modulation of $\chi^{(1)}(f)$ and its effect on the fundamental field $E(f)$.

In our experiment, we consider the SHG generated at the surface of a 310 nm layer of ITO by a probe beam at the fundamental frequency $f = 230$ THz, p-polarised and impinging from air at 45° incidence angle, near the Berreman resonance (see Supplementary Figure 1 for linear properties). The interface undergoes a strong time modulation upon optical pumping by a pump pulse at 53° incidence angle ($f = 230$ THz, 225 fs FWHM). The relative arrival time between the probe and the pump pulse is controlled via a delay stage. As illustrated in Fig. 1(b), a spectrometer is then used to analyze the effect of the modulation on the reflected beam, both at the fundamental and harmonic frequencies, by using different spectral filters.

**Temporal modulation of SHG**

The ITO thin film exhibits an ENZ frequency of 248 THz: below this frequency, the ITO layer is metallic and reflective, while above this frequency it is dielectric and transparent. As the ENZ frequency is expected to decrease under infrared pumping due to intraband transitions [20], we choose a probe frequency of 230 THz, in the metallic region. In this way, a large reflectivity drop can be observed during the metallic to dielectric transition under optical pumping. We can also directly measure the linear changes of the thin film at $2f$ as a reference, by using the SHG generated by a beta barium borate crystal as a probe beam.

We focus our study on the second harmonic generation in reflection as: 1) SHG in a centrosymmetric material, such as ITO, only comes from a relatively thin region close to the surface and 2) most of the reflected beam at the fundamental frequency comes from the first air/ITO interface, due to the strong absorption of the Berreman mode at the operating frequencies leading to little second harmonic generation from the second ITO/substrate interface. This configuration allows to model the experiment with the SHG produced at a time-varying interface and simplifies interpretation of the results, as the effect of pulse propagation and its thickness-dependent self- and cross-phase modulation can be neglected in the first approximation. We also choose a low probe intensity ($< 5$ GW/cm$^2$) so that the generation of second harmonic in the absence of the pump is non-perturbative, and time-modulation comes only from the much stronger pump beam (see Supplementary Figure 2 for power dependence of SHG).

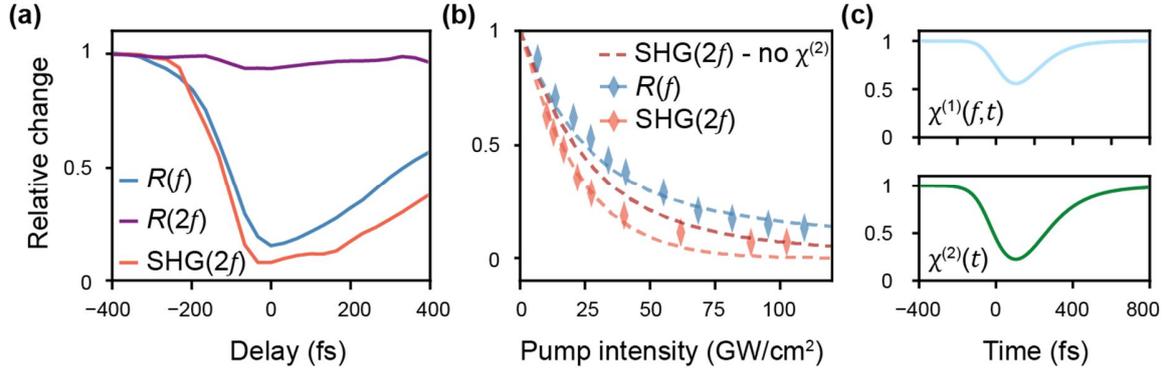

**Figure 2.** Comparison of the modulation amplitude of fundamental and harmonic signals. **(a)** Relative change in the total intensity of the reflected signal (integrated spectrum) as a function of delay, for the probe beam at the fundamental frequency $f$ (blue curve), its second harmonic at $2f$ (red curve) and another probe beam at $2f$ (purple curve), for a pump intensity of 100 GW/cm². **(b)** Pump intensity dependence of the maximal relative change in SHG (red diamonds) and reflected fundamental at $f$ (blue diamonds). The agreement with the anharmonic oscillator model (dashed lines) is excellent for the fundamental, as well as for the harmonic at low power, when taking into account the variation in $\chi^{(2)}$ (dark red dashed line for a model without $\chi^{(2)}$ changes). **(c)** Predicted relative change in susceptibility $\chi^{(1)}(f)$ (top, light blue) and $\chi^{(2)}$ (bottom, dark green) as a function of time for a pump intensity of 100 GW/cm² in a 310 nm film of ITO at 230 THz.

We excite the medium with a short laser pulse and use high pump powers to approximate a step-like variation of the reflectivity, whose sign is opposite to what previously observed in the time diffraction of an ITO/Gold film [17]. The modulation of the sample reflectivity at the fundamental frequency $R(f)$ is shown in Fig. 2(a) as a function of the relative pump/probe delay (continuous blue curve), for a pump intensity of 100 GW/cm². One can see that the modulation of the SHG (red curve) is larger than the modulation at the fundamental frequency. In contrast, the modulation experienced by a probe beam at $2f$, $R(2f)$ (purple curve), is 13 times smaller, and therefore the contribution of the temporal variation of the linear susceptibility $\chi^{(1)}(2f)$ to the SHG dynamics will be neglected in the following discussion (see Supplementary Figure 3 for more details).

At first sight, one could think that a decrease in the reflected signal at the fundamental frequency would lead to an increased SHG signal, as more field would penetrate the medium and contribute to the nonlinear polarization. However, this intuition does not account for the nature of the Berreman mode, which is a surface, absorptive mode. Near the resonance, a large electric field (mainly orthogonally polarized) exists just inside the ITO layer, which is driving the nonlinear polarization; when the medium becomes more dielectric this surface field becomes weaker (see Supplementary Figure 4), driving down the SHG signal as we see in Fig. 2(a).

In order to compare the time-varying effect at the fundamental and harmonic frequencies, we study the reflectivity modulation depth as a function of the pump intensity, as shown in Fig. 2(b). The difference between the modulation amplitude at the fundamental frequency (blue diamonds) and that of the SHG (red diamonds) is larger at low powers, below the saturation of the medium response, which clearly shows that time-varying effects are enhanced for SHG. The modulation depth also reaches a saturation earlier for the SHG in Fig. 2(b), as one would expect given the quadratic dependence of the nonlinear polarisation on the fundamental fields.

A simple Drude model of the ITO permittivity with a time-varying plasma frequency and electron scattering rate enables to calculate the expected change of the linear susceptibility $\chi^{(1)}(f,t)$ (see Methods). The calculated change in the Fresnel reflection coefficient of the interface as a function of pump intensity reproduces well the evolution of the modulation amplitude at the fundamental frequency (dashed blue curve). By simply using the square of the modulated fundamental field $E(f)$, as the source of the second-order polarisation $P^{(2)}(2f)$, one falls short of reproducing the measured SHG contrast (dashed red curve).

However, when adding the contribution of a time-varying $\chi^{(2)}(t)$, the agreement with the data becomes excellent (dashed orange curve). This is because the evolution of $\chi^{(1)}(f,t)$ and $\chi^{(2)}(t)$ are not independent, and modulating one always implies a modulation of the other. This is predicted by using an anharmonic oscillator model [28] (see Methods). Surprisingly, the model shows in Fig. 2(c) that the depth of the modulation of the nonlinear susceptibility $\chi^{(2)}(t)$ is expected to be almost double than that of the linear susceptibility $\chi^{(1)}(f,t)$. The large changes in the ITO's material properties (plasma frequency, electronic scattering rate) have a consequential imprint on the second order nonlinear susceptibility for the pump intensities used in this work.

**Time diffraction of SHG**

The experimental SHG spectrum carries evident signatures of the step-like temporal modulation induced by a 100 GW/cm² pump beam. As plotted in Fig. 3(a), the spectrum clearly broadens and redshifts at slightly negative delays, when the probe arrival time coincides with the pump leading edge, which corresponds to the excitation of the medium. As pointed out in previous studies of a time-varying mirror [5,23], frequency shift and broadening can be respectively linked to changes in the phase and amplitude of the complex reflection coefficient of the system. Although qualitatively similar to the time-diffraction at the fundamental frequency (see Supplementary Figure 5 for a detailed comparison), the overall shift of the harmonic spectrum is twice as large, with a much larger suppression of the peak around the unmodulated frequency. Both spectral broadening and shift increase with the magnitude of the time-modulation, driven by the pump power, although the effects saturate at about 70 GW/cm².

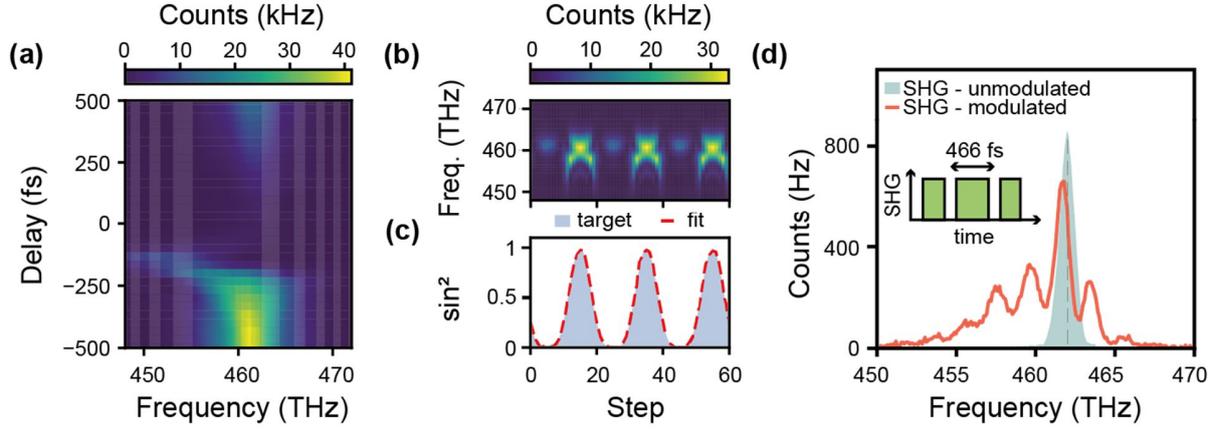

**Figure 3.** Spectral modulation of the harmonic signal. **(a)** SHG intensity spectra as a function of delay, for a probe duration of 225 fs and a pump intensity of 100 GW/cm$^2$. **(b)** Time-modulated SHG spectra against delay steps and **(c)** a ridge regression (red dashed curve) to link the step input to a target function (blue-shaded area). **(d)** SHG diffraction spectrum for two slits in time separated by 466 fs, for the same pump intensity of 33 GW/cm$^2$. The original probe pulse, centered at 462 THz with a width of 1.12 THz, is indicated with the green-shaded area.

Hence, the modulation of SHG signal in the non-perturbative regime is very susceptible to changes in pump-probe delay both in amplitude and frequency. This strong nonlinearity could have potential uses for optical computing and machine learning. As an example, a simple reconstruction algorithm can be used to build various waveforms such as square waves by simply sinusoidally varying the delay, as illustrated in Fig. 3(b) (more details in Supplementary Figure 6). The regression in Fig. 3(c) (red dashed curve) shows good agreement with the target shape, with a mean square error of 1.57×10$^{-4}$ for a sin$^2$ shape.

Finally, more complex temporal modulations of SHG are possible and lead to richer spectral dynamics. As an example, we performed a double-slit diffraction experiment, similar to the one recently reported in a different ITO sample [17]. The probe pulse is now stretched to 691 fs by using a 4f system and interacts with the medium modulated by two subsequent 225 fs pump pulses, the probe arrival time set in between the two slits' peak modulation. The reflected SHG spectrum is then measured. A clear broadening of the SHG spectrum (originally 1.12 THz FWHM, green shaded area) is observed in Fig. 3(d), together with the spectral oscillations characteristic of double-slit diffraction. The period of the oscillation scales inversely with the slit separation in time, in agreement with a simple Fourier theory of time diffraction (see Methods and Supplementary Figures 7 and 8). The visibility of the spectral modulation, defined as (max-min)/(max+min) for the first order peak, has a constant value of $0.58 \pm 0.01$ across the range of measured slit separations, comparable to the oscillations observed for the fundamental in a previous experiment[17]. As the visibility is a measure of the coherence of the time modulation, we can conclude that all the time-varying effects leading to SHG modulation, discussed in Fig. 1(a), combine in a coherent way, i.e. all new frequencies generated by the two slits at different points in time interfere coherently. Finally, thanks to the strong suppression of the original, unmodulated SHG spectrum in the time-modulated signal, one can also extract more information on the fast dynamic of ITO. For instance, the ratio in efficiency between the lower and higher frequency pulses which is linked to fast excitation and relaxation scales [17] is consistent with the presence of a short ∼ fs relaxation scale proposed by Lustig et al. [18] (see Methods and Supplementary Figure 8).

**Conclusion**

In conclusion, we have studied a perturbative process, SHG, in the presence of a non-perturbative modulation induced by optical intraband pumping of a time-varying ITO surface. By comparing a classic anharmonic oscillator model with experimental data, we have demonstrated an enhancement of the SHG modulation depth, which originates from a combination of the quadratic dependence of the nonlinear polarisation on the fundamental fields and a strong modulation of the nonlinear susceptibility, here reported for the first time. This enables ultrafast switching of signals in the visible range with high contrast, and we expect it to apply to other parametric processes like FWM and THG in ITO as well as in other time-varying materials, with potential applications in signal processing and optical communications. As a consequence of the enhanced effective nonlinearities, time-varying effects are also amplified in the frequency domain. The frequency broadening and shift of the harmonic spectrum is qualitatively similar but twice as large than that of the fundamental frequency. The strong sensitivity of the spectral features to the delay time between pump and probe can be exploited, for instance, to implement machine learning schemes. Finally, we have shown the potential of harmonic modulation, particularly of double-slit time diffraction, for ultrafast spectroscopy of time-varying media, where fine and reproducible spectral features can be used to reconstruct the medium complex time dynamics and to potentially reveal multiple timescales that are shorter than the probing pulses.

**Acknowledgements**

E.G. acknowledges support from the Simons Foundation (855344,EG).
R.S., S.V, JBP, SAM acknowledge support from UKRI (EP/V048880).
JBP acknowledges support from the Gordon and Betty More Foundation.
SAM acknowledges the Lee-Lucas Chair in Physics.

**Author Contributions Statement**

Conceptualization: RS, SV, RT, EG
Methodology: RT, SV, RS
Software: RT, SY, TVR, DS
Investigation: RT, SV, SY, DS, TVR
Visualization: RT, RS, SV, SY, DS, TVR
Funding acquisition: RS, JBP, SAM, SV, EG

Project administration: RS, JBP, SAM
Supervision: RS, SV
Writing – original draft: RT, SV, RS
Writing – review & editing: RT, SV, RS, TVR, JBP, EG, SAM, SY

**Competing Interests Statement**

The authors declare that they have no competing interests.

**Methods**

**The anharmonic oscillator model**

The following derivation is reproduced from Robert Boyd's *Nonlinear Optics* [28] where it can be found in full length. The electron position $x$ follows the equation of motion:

$$\ddot{x} + 2\gamma\dot{x} + \omega_0^2 x + ax^2 = -\frac{e}{m}E(t)$$

where $\omega_0$ and $\gamma$ are respectively the resonant frequency and the damping term, $-e$ is the electron charge, $m$ the electron mass and $a$ a nonlinear coefficient. Assuming the nonlinear term is weak in comparison to the restoration force, one can then derive the linear susceptibility of a single Drude-Lorentz oscillator using a perturbation expansion:

$$\chi^{(1)}(\omega) = -\frac{\omega_p^2}{D(\omega)}$$

where $\omega_p$ is the plasma frequency of the material and $D(\omega)$ the complex denominator of the Lorentz model:

$$D(\omega) = \omega^2 - \omega_0^2 + i\gamma\omega$$

Then, the second order nonlinear susceptibility can be derived to be:

$$\chi^{(2)}(2\omega, \omega, \omega) = \frac{ea}{m}\frac{\omega_p^2}{D(2\omega)D(\omega)^2}$$

ITO follows a Drude dispersion of the form:

$$\varepsilon_r(\omega) = \varepsilon_\infty - \frac{\omega_p^2}{\omega^2 + i\gamma\omega}$$

which matches with the Drude-Lorentz model in the case of $\omega_0 = 0$ i.e. the absence of restoring force on the electron, which is expected when considering only carriers in the conduction band as in the Drude model. Note that the Drude-Lorentz model assumes a single oscillator, which leads to the absence of an $\varepsilon_\infty$ high-frequency permittivity constant in the model. Nevertheless, the anharmonic model informs us that the second order nonlinear susceptibility is expected to vary if the Drude parameters $\omega_p$ and $\gamma$ vary following:

$$\chi^{(2)}(2\omega,\omega,\omega) = \frac{ea}{m} \frac{\omega_p{}^2}{(4\omega^2 + 2i\gamma\omega)(\omega^2 + i\gamma\omega)^2}$$

At high intensities the model is expected to fail due to higher order processes that are not accounted for [29], but its agreement with experimental data at low intensities is remarkable.

To model the change in SHG and other fundamental signals as a function of delay and intensity, the single slit aperture function is modelled as:

$$f(t) = \frac{1}{(1 + e^{-\alpha t})(1 + e^{\beta t})}$$

where $\alpha$ and $\beta$ characterize the dynamics of the excitation and relaxation of the modulation. The material properties are then expressed as

$$\omega_p(t, I) = \omega_p{}^{(0)}(1 + a \times I)f(t)$$
$$\gamma(t, I) = \gamma^{(0)}(1 + b \times I)f(t)$$

Where $f(t)$ has been normalized, $a$ and $b$ are coefficients fitted from experimental data, $I$ is the illuminating pulse peak intensity, $\omega_p{}^{(0)}$ and $\gamma^{(0)}$ the plasma frequency and electron scattering coefficient at rest and $\omega_p(t, I)$ and $\gamma(t, I)$ their modulated counterparts. Using a transfer matrix method, we can then compute the reflection of the fundamental, the surface field intensity at the origin of the second harmonic as a function of time as well as the evolution of the nonlinear susceptibility. This all allows us to compute the relative change in SHG signal.

In our experimental configuration, pump intensities of the order of 10s of GW/cm² are sufficient to induce large spectral broadening on a probe beam at the fundamental frequency, suggesting a speeding up of the medium time response as described in previous works [5]. Because of this steepening of the response at high pump power, it is possible to talk about time-diffraction and we model the system with a fast rise time of 4.36 fs (10-90%) and a slow decay time of 615 fs.

**Experimental setup**

The experimental setup is similar to that shown in our previous work [17]. Degenerate pump-probe experiments are led with 225 fs pulses with tunable frequency. Beam sizes were measured using a CCD camera and compared to knife-edge measurements. Delay between pumps and probe are controlled via 2 delay stages on the probe and one of the pumps. Pumps are incident at 6° angle difference on either side of the probe. Linear characterization of the sample was done using a broadband lamp and comparing the reflected signal to that of a reference silver mirror. For the double slit experiment, in order to experience the effects of both pump excitations the probe is broadened in time using a 4-f spectral filtering system. Reflected signals are sent to visible and near-infrared spectrometers. The non-degenerate measurement (pumping at frequency $f$, probing at $2f$) was done by placing a beta barium borate crystal at the focus of a telescope in the probe beam path and filtering out the fundamental frequency from the probe using a bandpass filter.

**The Fourier model**

A Fourier transform model [17] is used to inform us on the dependence of the diffraction orders peak intensity on the dynamics of the medium. The reflection coefficient amplitude of the ITO layer is expressed as

$$r(t) = A_1 \times f\left(t - \frac{S}{2}\right) + A_2 \times f\left(t + \frac{S}{2}\right)$$

where $S$ is the slit separation in time, and $A_1$ and $A_2$ are negative amplitude constants fitted from experimental data. The reflected spectrum is then expressed as $\text{FT}[r(t) \times E_{\text{probe}}(t)]$ where $E_{\text{probe}}(t)$ is the probe pulse electric field in time.

**Data and materials availability**

Source data are available for this paper and are deposited in a public repository (future link). All other data that support the plots within this paper and other findings of this study are available from the corresponding author upon request.

**Supplementary Information**

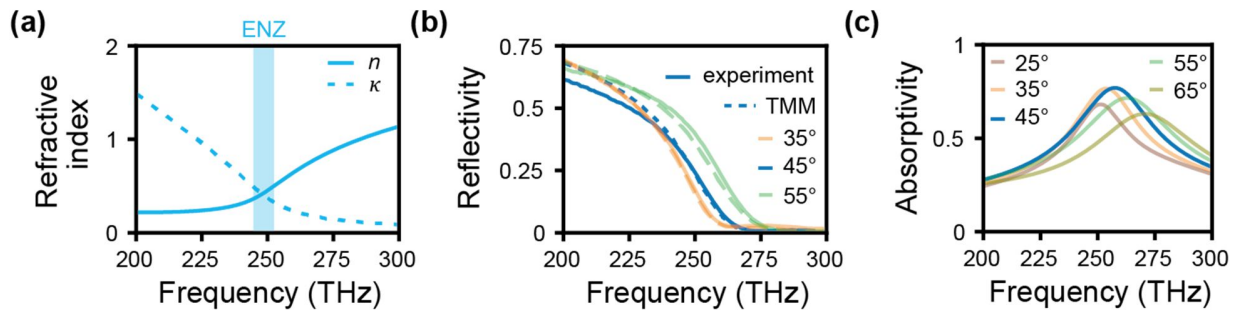

**Supplementary Figure 1**. Linear properties of the 310 nm ITO sample. **(a)** Refractive index of the ITO layer as a function of frequency, real (continuous) and imaginary (dashed curve). The ENZ region, centered at 248 THz, is highlighted with the shaded area. **(b)** Measured (continuous) and numerically computed with TMM (dashed) reflectivity spectrum for various incidence angles. **(c)** Corresponding simulated absorption spectra, showing a peak in absorptivity at 45 degrees, indicating the central frequency of the Berreman resonance.

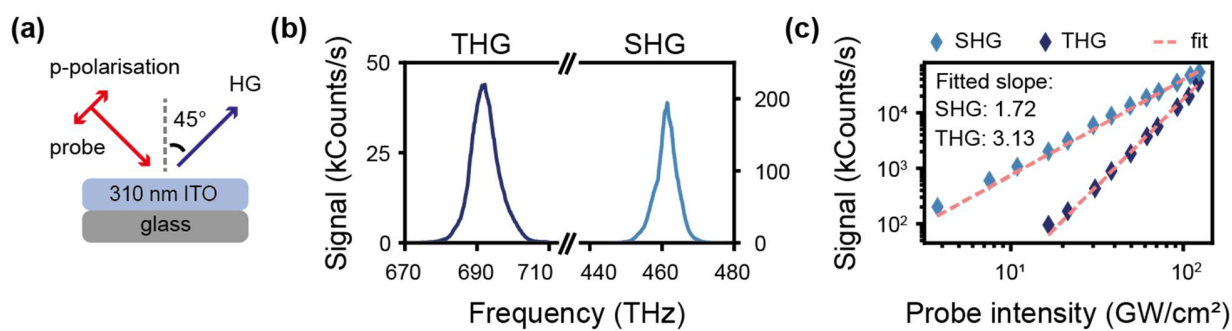

**Supplementary Figure 2**. Nonlinear properties of the 310 nm ITO sample. **(a)** Schematic of the harmonic generation measurement. **(b)** Third (left, dark blue curve) and second (right, light blue curve) harmonic signals for a carrier frequency of 230 THz and an illuminating intensity of 31 GW/cm$^2$. **(c)** Intensity dependence of the THG (dark blue diamonds) and SHG (light blue diamonds) signals on a logarithmic scale. Power law fits are shown in dashed pink.

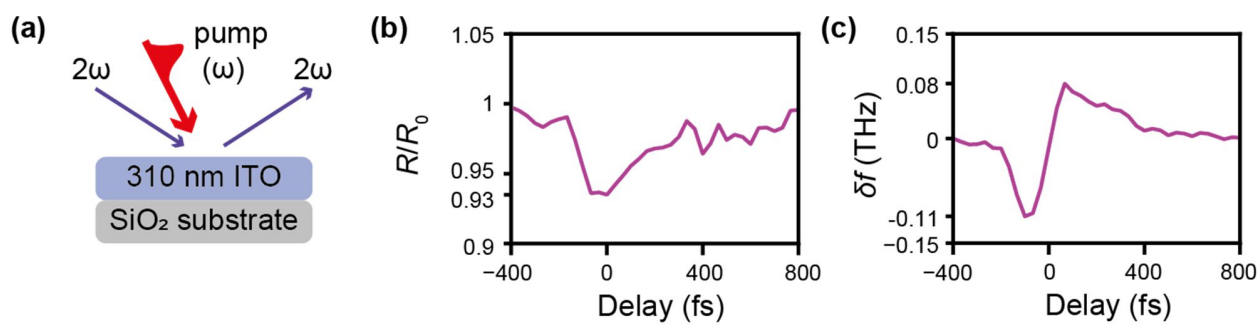

**Supplementary Figure 3**. Pump-probe experiment with a probe fundamental at frequency $2f = 460$ THz. **(a)** Diagram of the pump-probe measurement with a probe centered at 460 THz. **(b)** Relative reflectivity change as a function of delay. **(c)** Frequency shift of the probe spectrum as a function of delay.

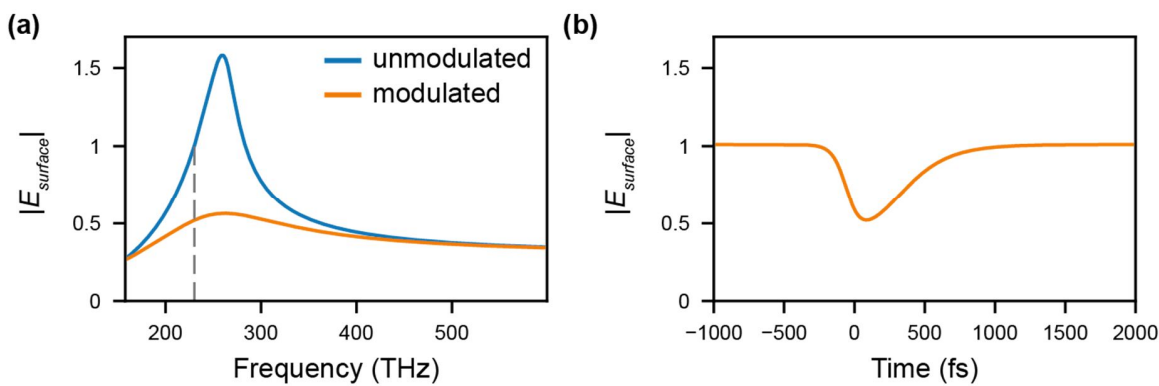

**Supplementary Figure 4**. Surface field at the air/ITO interface. **(a)** Simulated surface field amplitude spectrum, unmodulated (blue curve) and modulated with a pump at 100 GW/cm$^2$

(orange curve). **(b)** Simulated surface field amplitude as a function of delay for a pump intensity of 100 GW/cm$^2$ and a probe carrier frequency of 230 THz.

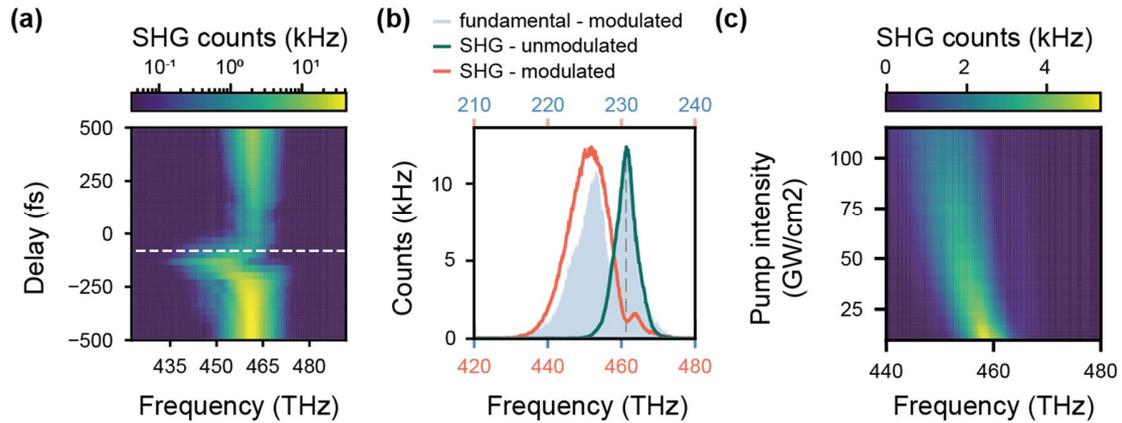

**Supplementary Figure 5.** Spectral modulation of the fundamental and harmonic signals. **(a)** Logarithmic scale SHG intensity spectra as a function of delay as shown in Fig. 3(a), for a probe duration of 225 fs and a pump intensity of 100 GW/cm$^2$. The white dashed line indicates the delay at which the spectrum in panel (b) is shown. **(b)** Comparison between the modulated fundamental (blue-shaded area) and SHG (red curve) spectra for a delay of -66 fs. The unmodulated SHG spectrum is shown in dark green for reference. The grey dashed line indicates the carrier frequency of the unmodulated fundamental/second harmonic. **(c)** Evolution of the SHG spectrum as a function of pump intensity, for a delay of -66 fs.

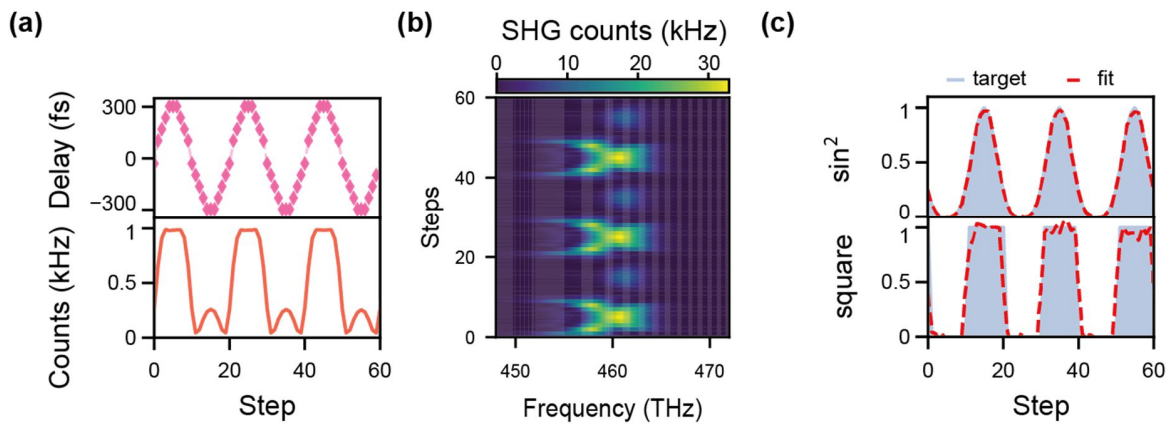

**Supplementary Figure 5.** Ridge regression on analogue time-modulated data. **(a)** Input of the regression algorithm: the delay steps are varied in a triangular pattern (pink curve, top), which leads to a modulation of the second harmonic counts with said steps (orange curve, bottom). **(b)** The interferogram of the SHG spectrum against step is recorded (as shown in Fig. 3(b)), the data

is used for a ridge regression to link the step input to any **(c)** target function as a function of step (blue-shaded area). The regression (red dashed curve) shows good agreement with the target shape, with a mean square error of $1.57 \times 10^{-4}$ for a $\sin^2$ shape (top, as in Fig. 3(c)) and $2.77 \times 10^{-2}$ for a square shape (bottom).

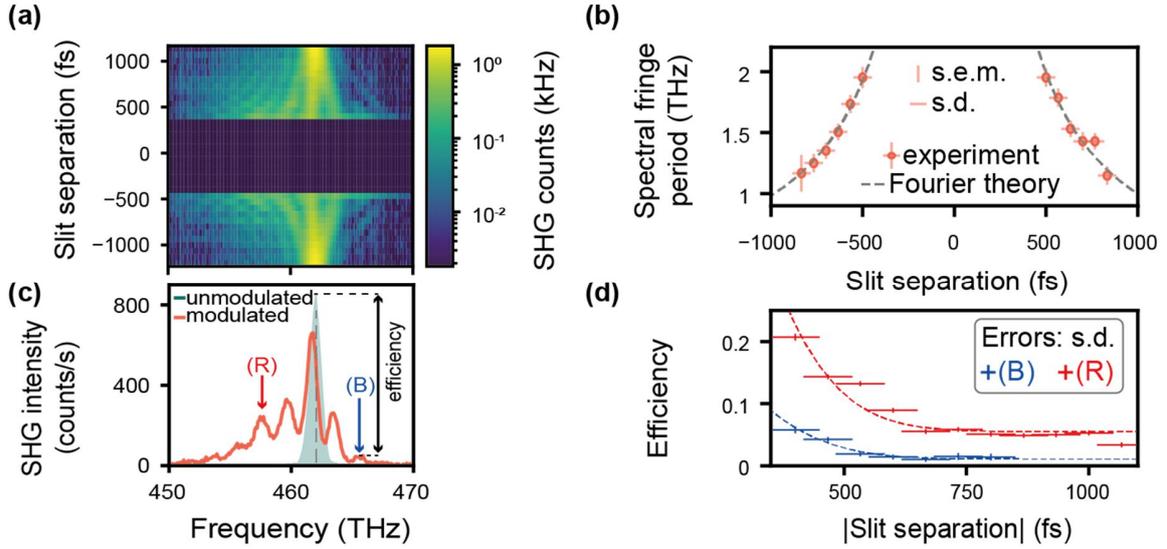

**Supplementary Figure 7.** A double slit experiment in time on second harmonic generated light from an ITO thin film. **(a)** SHG spectrum for various slit separations, on a logarithmic scale. Slit separations below 400 fs were not included as sum-frequency generation between two pump pulses becomes dominant. **(b)** Extracted period of the spectral oscillations against slit separation in time. The experiment (red circles) matches the inverse proportionality law from Fourier theory (dashed grey line, see Methods). The horizontal and vertical error bars represent the standard deviation (s.d.) and standard error of the mean (s.e.m.), respectively. **(c)** SHG diffraction spectrum for two slits in time separated by 467 fs as shown in Fig. 3(f), for the same pump intensity of 33 GW/cm². The original probe pulse, centered at 462 THz with a width of 1.12 THz, is indicated with the green-shaded area. **(d)** Efficiency of the 2$^{nd}$ order red peak (R) and blue peak (B) as a function of the slit separation. Frequencies generated far from the central peak carry information on the short timescale of the modulation. The existence of a 2$^{nd}$ blue-shifted peak is evidence of a short component in the medium recovery time. Errors are expressed as standard deviations. The trend is well fitted by the envelope of the probe Gaussian pulse (691 fs FWHM, dashed lines).

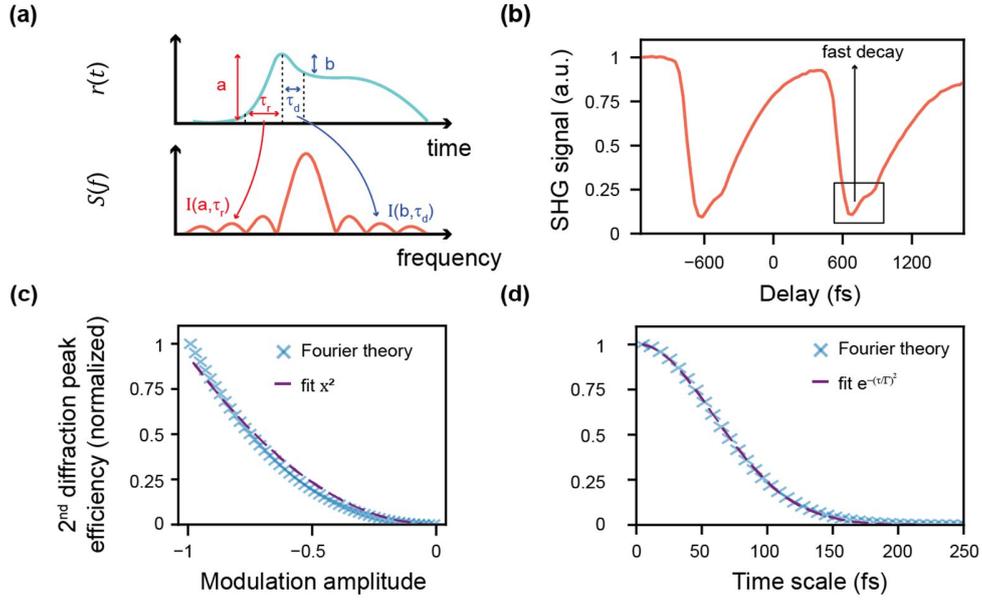

**Supplementary Figure 8.** Generated frequencies' dependence on dynamics and amplitude of changes in the medium. **(a)** Sketch of the respective effects of the fast excitation and relaxation time scale and amplitudes of the complex reflection coefficient r(t) on the diffraction order intensities of the spectrum S(f). **(b)** Reflected SHG signal measured as a function of delay for a double slit modulation (pump power 62.5 GW/cm$^2$, slit separation 1200 fs), with a probe duration of 225 fs to resolve the slit's individual effect on the SHG. **(c)** Modelled 2$^{nd}$ order diffraction peak dependence on modulation amplitude (blue crosses). To illustrate the quadratic behavior of the curve, a fitted curve is shown in dashed purple. **(d)** Modelled dependence of the 2$^{nd}$ order diffraction peak intensity on the red/blue side of the spectrum on the excitation/relaxation time (blue crosses). The agreement with a Gaussian fit (dashed purple curve) is excellent.

**Estimation of a fast relaxation time**

The Fourier model model is used to show that the 2$^{nd}$ and higher order diffraction peaks intensity on the red/blue side respectively depend on the excitation/relaxation times as well as the amplitude in reflection coefficient of said excitation/relaxation. As can be seen in Supplementary Figure 8, the behavior of the intensity of the 2$^{nd}$ order diffraction peak is quadratic with modulation amplitude and Gaussian (centered at 0 fs) with time scale. Writing down $a$ and $b$ as the respective amplitudes of the fast excitation and relaxation, and $\tau_r$ and $\tau_d$ their respective time scales, we can predict the relative intensities of the 2$^{nd}$ order diffraction peaks on the red and blue sides of the spectrum $I_{\text{red}}(a, \tau_r)$ and $I_{\text{blue}}(b, \tau_d)$ using the Fourier model. From our experimental data (see Supplementary Figure 7), we can extract $a$ and $b$ (see Supplementary Figure 6), the excitation time $\tau_r \sim$ 1-10 fs from the spectral extent of the red oscillations [17], as well as the ratio $I_{\text{blue}}(b, \tau_d)/I_{\text{red}}(a, \tau_r)$. We can then estimate a value for $\tau_d$ from the Fourier model. As we measure $a = 0.95$ and $b = 0.36$, we compute $I_{\text{blue}}(b, \tau_d)/I_{\text{red}}(a, \tau_r) = 0.14$ for $\tau_d = \tau_r$. From our experimental data, we measured an average value of $I_{\text{blue}}/I_{\text{red}} = 0.16 \pm 0.02$, hence the simulations are in agreement with the experiment. Note that the Fourier model is limited as it does not include effects such as dispersion of the sample or phase shifting of the probe, and thus can only give a qualitative understanding of the dynamics of the medium.